# A geometric theory for the origin and motion of sand dunes

Daan Beelen

*Utrecht University, Department of Physical Geography, Utrecht, the Netherlands*

**ABSTRACT**

In this study, a new theory for the spontaneous formation of sand dunes and related bedforms is proposed. The theory is based on the concept that smaller accumulations of sediment outpace larger ones due to differences in surface-to-volume ratio. From this geometric principle, it follows algebraically that for any nonzero bedload transport, the bed (sediment-covered surface) must be tilted upward along the current's direction, forming ripples and dunes. These ideas are validated by making the first-ever accurate predictions of dune speeds (migration rates) that are derived solely from measurements of stoss side (s) and lee face length (l). This prediction has very high statistical significance: $R^2= 0.92$, $P = 2.0\text{e-}122$, $n=250$. The geometric relationships defined here also predict the backwards migration of antidunes and explain the 'horned' barchan dune and mirrored parabolic dune shapes.

Dunes are amalgamations of sediment that migrate under the influence of a current. A common environment for dune formation is a desert, where sand is pushed into various shapes of dune that migrate in the dominant wind direction. Time-averaged dune velocities (migration rates) can reach up to 30 cm per day, sometimes affecting people and infrastructure (Liu et al., 2005). Outside of deserts, dunes are ubiquitous in fluvial, coastal, and marine sedimentary environments, and play a key role in desertification, river dynamics and coastal protection against sea level rise (Ruessink et al., 2002). A large portion of the sedimentary rock record is comprised of stacked fossil dunes (i.e. cross strata), which are used to reconstruct ancient current dynamics, paleo water depths and other paleoenvironmental conditions (Eastwood et al., 2012). Understanding dunes is therefore of major importance to a wide range of geological work from geohazard management to paleoenvironmental reconstruction and process sedimentology and stratigraphy (Kocurek, 1991, Rodríguez-López et al., 2014).

Explaining how and why dunes form is a historic research pursuit with the earliest ideas going back to essential concepts of incipient motion and sediment dynamics (Cornish, 1897, Einstein, 1950). Subsequent insights involve the convergence and divergence of fluids across an uneven bed (Kennedy, 1969) and the spontaneous inception of fluid turbulence (Nishimori and Ouchi, 1993). More recent theories explore the effects of random turbulent fluctuations and boundary layer profiles that shape and transport moving sediment into geometric shapes (Tsoar, 2001, Charru et al., 2013). Although many of these theories have been shown to be effective in numerically and experimentally explaining some of the observed dynamics of ripples and dunes (Engelund and Fredsoe, 1982), various problems persist. For example, theories such as those proposed in Liu (1957), Wiliams and Kemp (1979) or Charru et al. (2013) do not give a complete explanation as to why such regular and elegant landforms, as observed in nature, can arise from turbulent dynamics (Andreotti et al., 2002). Although this uncertainty can be partially mitigated by invoking concepts like morphodynamic self-organization, the emergence of very similar bedforms across environments with widely different levels of fluid turbulence, and even those approaching laminar conditions, remains difficult if not impossible to explain under the existing paradigms (Coco and Murray, 2007, Falqués et al., 2008, Baas et al., 2011, Coleman and Eling, 2000). Physical experimentation has also revealed that invoking obstacles to account for the initial formation and development of the observed phenomena is not essential and that bedform dynamics in clayey substrates is analogous to those in sand, despite theories predicting rather different interactions under such widely different sedimentological conditions (Maun, 2009, Schieber and Southard, 2009,

Schieber, 2011, Yawar and Schieber, 2017).

To address these issues, this paper explores an entirely new concept that lacks an 'a-priori' need for obstacles, turbulence, and complex paradigms such as turbulence or morphodynamic self-organization. Instead, the theory presented here is grounded solely in a limited set of geometric and algebraic axioms. In this paper, this theory is first outlined and then validated using measurements of the migration rates, shapes, and sizes of wind-driven barchan dunes in various settings. Although the theories presented here are mostly scale and shape invariant, large (15-300 m long) barchan dunes are the main bedforms used for this study's validation. The reason for this is that these dunes are large enough to allow for accurate measurements of their size and migration rate from satellite images, while their isolated arrangement ensured that their shapes and sizes are not affected by collisions and wind interference. Barchan dunes in close proximity also tend to vary widely in size and shape, which is also an important requirement for the validations presented here.

**Surface-to-volume ratios**

The surface-to-volume ratio (A/V) is the amount of surface area (A) compared to volume (V) of an object. Consider that the size of an object can be linked to a characteristic length (L), this is for example the length of a cube or the radius of a sphere. As an object's surface area increases quadratically to this length, the objects' volume increases to the third power. With increasing size, volume always increases more than surface area so larger objects always have a lower A/V compared to shape-equivalent, smaller objects (Fig. 1). This concept is long understood to play a fundamental role in sediment dynamics, as Stokes's law of sediment settling essentially describes a surface-to-volume ratio relationship between gravity and fluid drag, which relates linearly to a particle's settling rate in a fluid (Stokes, 1850). However, the implications of this key concept has yet to be explored explicitly in the context of bedforms.

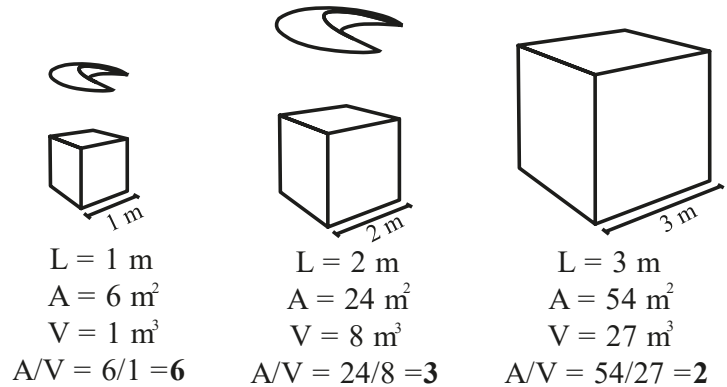
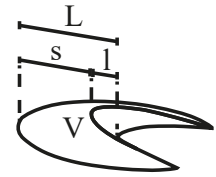

| | | |
|---|---|---|
| L = 1 m | L = 2 m | L = 3 m |
| A = 6 m² | A = 24 m² | A = 54 m² |
| V = 1 m³ | V = 8 m³ | V = 27 m³ |
| A/V = 6/1 = **6** | A/V = 24/8 = **3** | A/V = 54/27 = **2** |

*Figure 1. Schematic diagram showing the relationship between an object's size and its surface-to-volume ratio. In the case of a cube, the surface-to-volume ratio halves when the characteristic length of the cube doubles. This principle holds true for any object while the exact relationships change with shape.*

## THEORY

### Surface-to-volume ratios and sediment motion

As a current passes over a sediment-laden substrate (bed), a shear stress is exerted which may move the sediment forward if sufficiently strong. Some essential variables for bedload transport under a current therefore include shear stress, and a sediment volume (V) which has some sediment exposed to the current (A)[Footnote 1]. The moving sediment now has a surface-to-volume-ratio that is directly proportional to the amount of exposed, '*active*' sediment with respect to the amount of unexposed, '*passive*' sediment that is sheltered from the current. If we consider the work done by the current to be equal across the bed, then a doubling of the surface-to-volume ratio constitutes a doubling of sediment motion. We can formulate this statement as follows:

$$u \propto \frac{A}{V}$$

This statement forms the central axiom for the theories described here.

Aside from the surface-to-volume ratio, sediment migration rate is further controlled by other factors (e.g. current velocity) that combine to describe the constant of proportionality in this equation, so that:

*Equation* (1.1):
$$u = b * \frac{A}{V}$$

We refer to this proportionality as the 'scale-adjusted migration rate' (b) which is defined as:

$$b = u * \frac{V}{A}$$

Scale-adjusted migration rates exist for any object moving under a current and depend on environmental factors related to sediment motion like shear stress, current speed, fluid density, sediment density, water table depth etc. While surface-to-volume ratios can vary widely for adjacent dunes or sediment accumulations, scale-adjusted migration rates should be roughly constant for adjacent dunes in a similar environment.

To quantify these geometries, let's define the length facing the current (stoss side) as 's', and the length facing away from the current (lee side) as 'l'. The sum of these represent the total length of the sediment volume parallel to the current direction denoted by 'L' (Fig 2). For the width of the exposed surface we can use 'w'. Therefore, the surface area exposed to the current (A) is approximately [Footnote 2]:

$$A = \sqrt{h^2 + s^2} * w$$

The partially exposed sediment volume is:

$$V = \frac{1}{2} h(s + l) * w$$

Therefore, the surface-to-volume ratio of the migrating sediment is:

$$\frac{A}{V} = \frac{\sqrt{h^2 + s^2} * w}{\frac{1}{2} h(s + l) * w}$$

The width cancels out so that:

$$\text{Equation (2):} \quad \frac{A'}{V'} = \frac{2\sqrt{h^2 + s^2}}{h(s + l)}$$

Cancelling out widths alters the dimensions for these geometries within the current-parallel perspective: Exposed surface area (A') is now one-dimensional and volume (V') is two-dimensional. Although counter-intuitive, this operation is algebraically valid and shows how the relationship described in equation (1.1) is uniform from a current-perpendicular (i.e. 90 degrees from current-parallel) perspective. This notion is important as it helps explains the commonly observed tendency for sediment to form parallel ridges (i.e. ripples or transverse dunes) that move uniformly, perpendicular to the direction of the current.

To discover why the bed must tilt under the forced exerted by a current, let's revisit the definition for scale-adjusted migration rate but now in the 2D perspective:

$$b = u * \frac{V'}{A'}$$

This is:

$$b = u * L * \frac{h}{2A'}$$

The key here is that surface-to-volume ratio can be redefined in terms of the gradient of the windward stoss side: Angle β:

$$b = u * L * \frac{1}{2} \sin \beta$$

Or alternatively:

$$\text{Equation (1.2):} \quad u = \frac{2b}{L * \sin \beta}$$

This equation demonstrates that under the initial statement (equation 1) it is necessary for the to be bed to become tilted at a nonzero angle β when the sediment is moved.

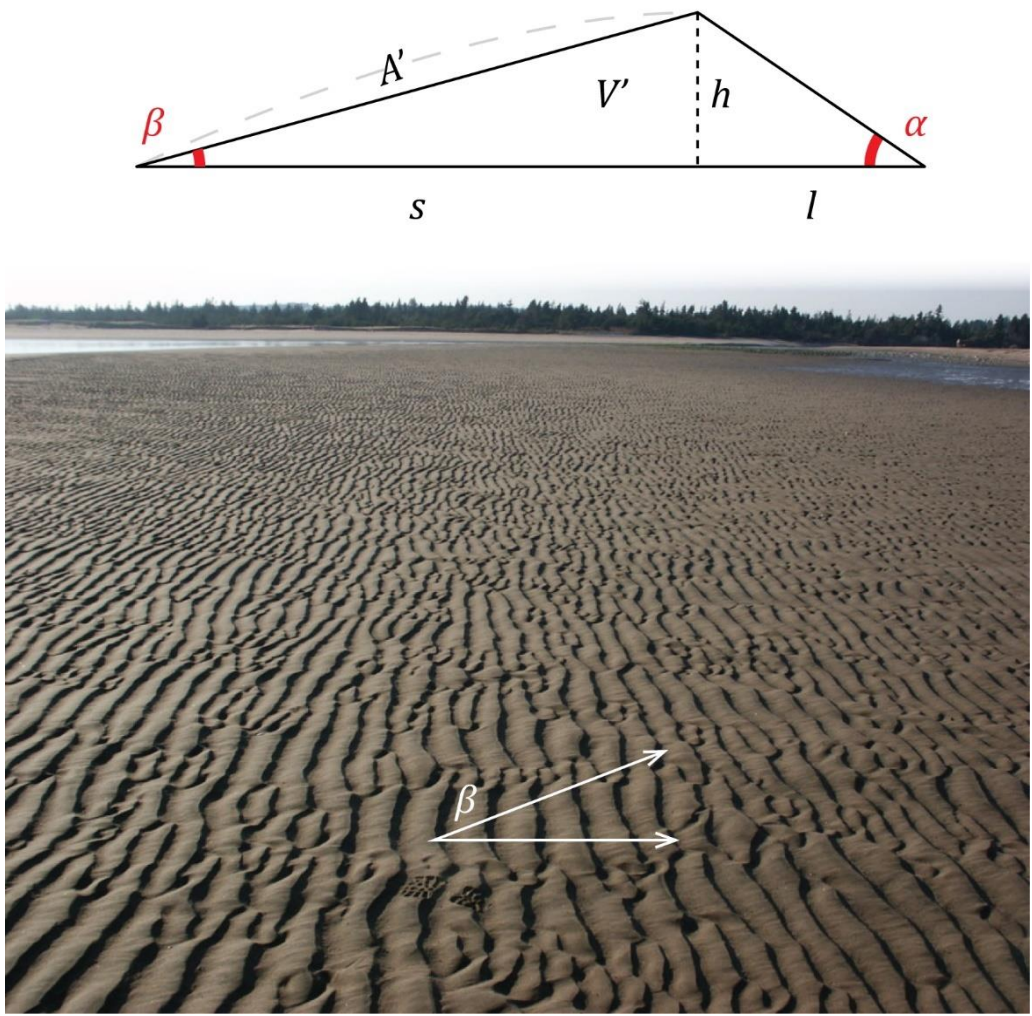

*Figure 2. Simplification of an equilibrium-condition bedform in the wind-parallel cross section. The stoss side has base s and height h with sharp angle β which raises the dune from the bed. The lee triangle has base l and height h with sharp angle equal to the angle of repose: α. Dune width is not shown in the diagram but is denoted by w. Diagram is accompanied with a photograph of a rippled surface that demonstrates that an upwards tilting at angle β explains this naturally-occurring geometry.*

**Sediment transport**

Since we are exploring the transport of sediment, it is useful to quantify the amount of sediment that is moved past a point in space, per unit of time. This variable is proportional to the volume and the speed of the moving sediment. Considering that longer objects move less volume past an imaginary point per unit of time, the transport is also inversely proportional to the object's length. In current-parallel cross section, we can thus define sediment transport as:

$$Q' = \frac{uV'}{L}$$

This is equal to:

$$Q' = \frac{uV'}{L} = u\frac{\frac{1}{2}hL}{L}$$

Length cancels out so:

$$Q' = \frac{1}{2}uh$$

Consider again that:

$$b' = u * \frac{V'}{A'}$$

This is equal to:

$$b' = u * \frac{\frac{1}{2}hL}{A'} = \frac{1}{2}uh * \frac{L}{A'}$$

Therefore:

$$b' = Q' * \frac{L}{A'}$$

Note the similarity between scale-adjusted migration rate and sediment transport, which have the same constituents of migration rate and height. Scale-adjusted migration rate and sediment transport therefore have a similar meaning and will always vary in tandem but can never be exactly the same [Footnote 3]. Since all geometries are non-negative, we can derive the following dimensionless relationship:

*Equation* (3.1):
$$\frac{b' - Q'}{Q'} = \frac{L - A'}{A'}$$

To appreciate what this equation means, let's focus on the right-hand side term. This term implies a relative reduction of the surface area exposed to the current (A') across the length of the sediment accumulation (L). Such a reduction of surface area entails a contraction of the bed in the direction of the current which we can call 'bed shortening'. We can expand this term to find:

$$\frac{L - A'}{A'} = \frac{l + s}{A'} - \frac{A'}{A'} = \frac{l}{A'} + \frac{s}{A'} - 1$$

Note that h and l relate to each other only in terms of the angle α:

$$h = l \tan \alpha$$

Furthermore:

$$\frac{h}{A'} = \sin \beta$$

And:

$$\frac{s}{A'} = \cos \beta$$

Again, we can redefine this in terms of stoss side angle β and lee side angle α:

$$\frac{L - A'}{A'} = \frac{\sin \beta}{\tan \alpha} + \cos \beta - 1$$

Therefore:

*Equation* (3.2): $$\frac{b' - Q'}{Q'} = \frac{\sin \beta}{\tan \alpha} + \cos \beta - 1$$

Equation (3.3) represents an equilibrium condition whereby the sediment will tend towards a nonzero steepness of the windward size of the sediment (angle β) with respect to horizontal. Therefore, whenever Q>0, the sediment must be raised from the bed at angle β, which constitutes the formation of a bedform. It is important to realize here that equations (1 and 2) apply equally to fully formed bedforms and any sediment-laden bed or sediment accumulation. In contrast, equations (3) define an equilibrium condition that will be satisfied given enough time for bedforms to form and develop their equilibrium geometry.

**Explaining barchan dune shapes**

To investigate how equation (1) manifests from the perspective of a single, isolated parcel of sediment. we can apply the aforementioned principles to a sediment accumulation that has not been fully developed into a bedform (e.g. a protodune *sensu* Qian et al. 2021), [Footnote 4]. This sediment accumulation can be dissected into equal portions in the wind-parallel direction. The segments at the fringes have higher surface-to-volume ratio and thus migration rate compared to the center, which cause the fringes to outpace the center and develop a 'horned' shape that is characteristic for barchan dunes (Fig. 3A). Moving from the center to the fringe, each subsequent portion decreases more in volume than in exposed surface area according to the inverse-square law. The surface-to-volume ratio therefore increases from the center towards the fringes in a power-law relationship, resulting in parabola-shaped horns (Fig. 3A). In this example, we started from a sediment mound, but starting with a fully formed barchan dune geometry gives the same result, meaning that any isolated sediment mound moved by current will develop a geometry with horns pointing away from the wind.

Secondly, we consider a sediment accumulation that is dissected vertically. Higher surface-to-volume at the top as compared to the bottom generates a parabola-shaped, windward stoss side and causes the dune to oversteepen on the forward facing lee face (Fig. 3B). As a dune forms, sediment that becomes oversteepened will cascade down the lee face that initially forms at the base and proceeds to develop upwards. Since this material cascades down under gravity and is not affected by the wind, the lee face angle α is always equal to the angle of repose. The upwards creeping lee face will eventually meet the parabolically-shaped stoss side nearby the dune's roughly horizontal top.

When both wind-parallel and vertical profiles are considered, the dune logically develops an geometry characterized by a parabola-shaped stoss side with a near-horizontal top, a parabola-shaped footprint surface area, and a diagonal, forward facing lee face with a slope that is equal the angle of repose of the sediment (angle α; Fig. 3C).

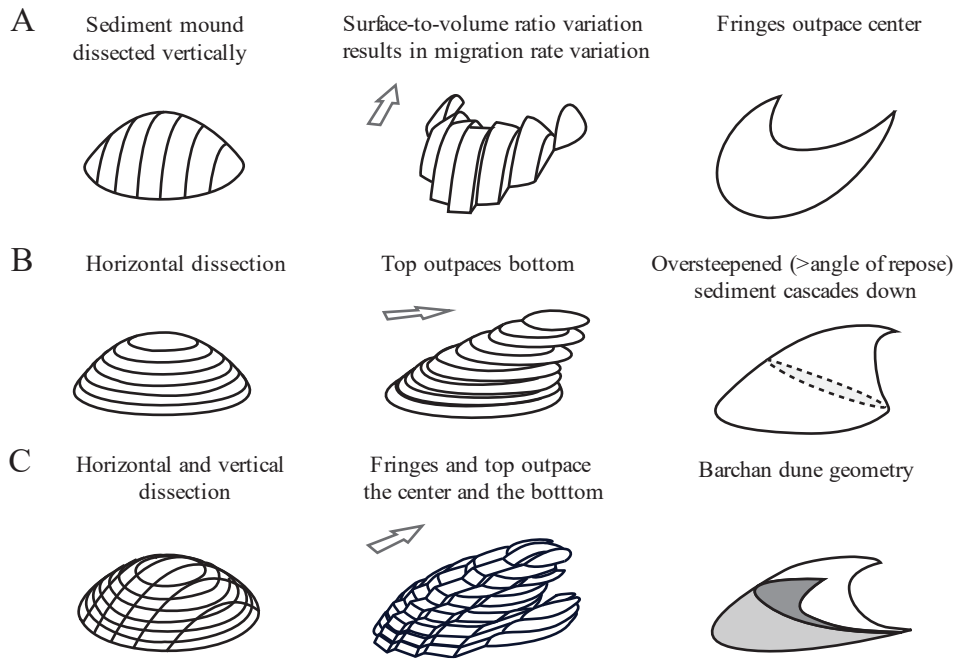

*Figure 3. Variation in surface-to-volume ratio explains barchan dune geometry. A. Consider a sediment mound that is dissected perpendicular to the wind direction.* In this illustration, the mound is dissected into seven volumes. The volumes at the fringes have a higher surface-to-volume ratio and correspondingly, higher migration rate. This implies that the fringes outpace the center leading to the characteristic 'horned' barchan dune shape. B. Dissections across the vertical imply a relatively fast top, leading to a parabolic stoss side and oversteepening with associated sediment cascading down the forward lee face. C. Dissections in both dimensions result in a complete barchan dune equilibrium shape with a parabolic stoss side, a diagonal lee face and a 'horned' footprint geometry. The arrow shows the dominant wind direction and direction of dune migration.

**Wind power**

Now that we explored the implications of the surface-to-volume ratio axiom in the perspectives of dunefields and individual dunes, let's apply it to a universal case. We can achieve this by defining wind power across the dune, which is the kinetic energy of the wind over time whereby the windspeed also determines the 'length' of the parcel of air that moves past the dune. In the wind-parallel cross section wind power is therefore equal to the height multiplied by the 'effective' windspeed cubed:

$$P'_w = \frac{1}{2} h u_w^3 * \rho_w$$

The power of the dune can be defined as the loss of gravitational potential energy of the sand that cascades down the lee face over time:

$$P'_d = h^2 u * g \rho_s$$

Since all the power of the dune comes from the wind we can state that:

$$P'_w = P'_d * C$$

Here, 'C' is the ratio of windpower converted into dune power. Therefore:

$$\frac{1}{2} h u_w^3 * \rho_w = h^2 u * g \rho_s * C$$

Which is:

Equation (4.1) $\qquad u_w^3 = Q' * 2gC \frac{\rho_s}{\rho_w}$

Sediment transport is therefore directly proportional to the effective windspeed. We can use this to determine a link between the amount of current energy in the system and the geometries and sizes of dunes. Consider equation (3.1).

$$u_w^3 = b' * \frac{A'}{L} * 2gC \frac{\rho_s}{\rho_w}$$

Now consider equations (1.2) and (3.2):

$$u_w^3 = \left(u * L * \frac{1}{2} \sin \beta\right) * \left(\frac{\sin \beta}{\tan \alpha} + \cos \beta\right)^{-1} * 2gC \frac{\rho_s}{\rho_w}$$

Which is:

$$u_w^3 = u * L * \frac{\sin\beta * \sin\alpha}{\sin(\alpha + \beta)} * 2gC\frac{\rho_s}{\rho_w}$$

Since dune length is inversely proportional to surface-to-volume ratio (Equation 1.1), we can eliminate both migration rate and dune length from the previous equation by defining a 'universal' scale-adjusted migration rate. Let's call this constant 'k'. This trick allows us to determine a direct link between effective windspeed and stoss side gradient:

$$Equation\ (4.2) \qquad u_w^3 = k * \frac{\sin\beta * \sin\alpha}{\sin(\alpha + \beta)} * 2gC\frac{\rho_s}{\rho_w}$$

## VALIDATION

To validate the theory, time series of actively migrating dunes are obtained from Landsat satellite images. These images are georeferenced and time-stamped, so the sizes and migration rates of dunes can be recorded and plotted, in a similar way to previous studies like in Liu et al. (2005), Michel et al. (2018) and Baird et al. (2019). Measurements of lee face and stoss side lengths are made to estimate exposed surface areas and dune lengths according to equation (2).

For each dunefield considered, a rectangular sampling perimeter is outlined, and all footprint surface areas (A), stoss side lengths (s) and lee face lengths (l) are measured. Dune displacement between timesteps are measured to calculate their migration rates. Timesteps between sequential satellite images are always several years (2-32 years) long, to accommodate seasonal changes in wind patterns and migration rates (Bristow, 2019). All measurements are in the supplementary information *(S1)*.

### Linear relationship between migration rate and surface-to-volume ratio

Measurements confirm the primary hypothesis put forth in this study, which is that for adjacent dunes, migration rates correlate linearly to surface-to-volume-ratios (Fig. 5).
It should be noted here that it has been reported widely in the scientific literature that dune migration rates are linearly proportional to the inverse of dune heights. This observation is attributed to Bagnold, (1941) and as such is sometimes referred to as 'Bagnold's Law' (Liu et al., 2005). Indeed, we can state that the inverse of any one-dimensional geometry of the dune correlates linearly to the surface-to-volume ratio and consequently, its migration rate. To explain this, consider:

$$A \propto L^2$$

$$V \propto L^3$$

$$\frac{A}{V} \propto \frac{L^2}{L^3} = L^{-1}$$

Therefore:

$$u \propto \frac{A}{V} \propto L^{-1} \propto h^{-1}\ etc.$$

Even the inverse of a dune's circumference (also a one-dimensional parameter) correlates linearly to its migration rate. When considering a two-dimensional geometry, like footprint surface area or exposed surface area, consider that:

$$V = A^{\frac{3}{2}}$$

So:

$$\frac{A}{V} \propto \frac{A}{A^{\frac{3}{2}}} = A^{-\frac{1}{2}}$$

So:

$$\frac{A}{V} \propto A^{-\frac{1}{2}} = \frac{1}{\sqrt{A}}$$

Dune migration rates are therefore inversely correlated to the square root of the surface area (Fig. 5):

$$u \propto \frac{1}{\sqrt{A}}$$

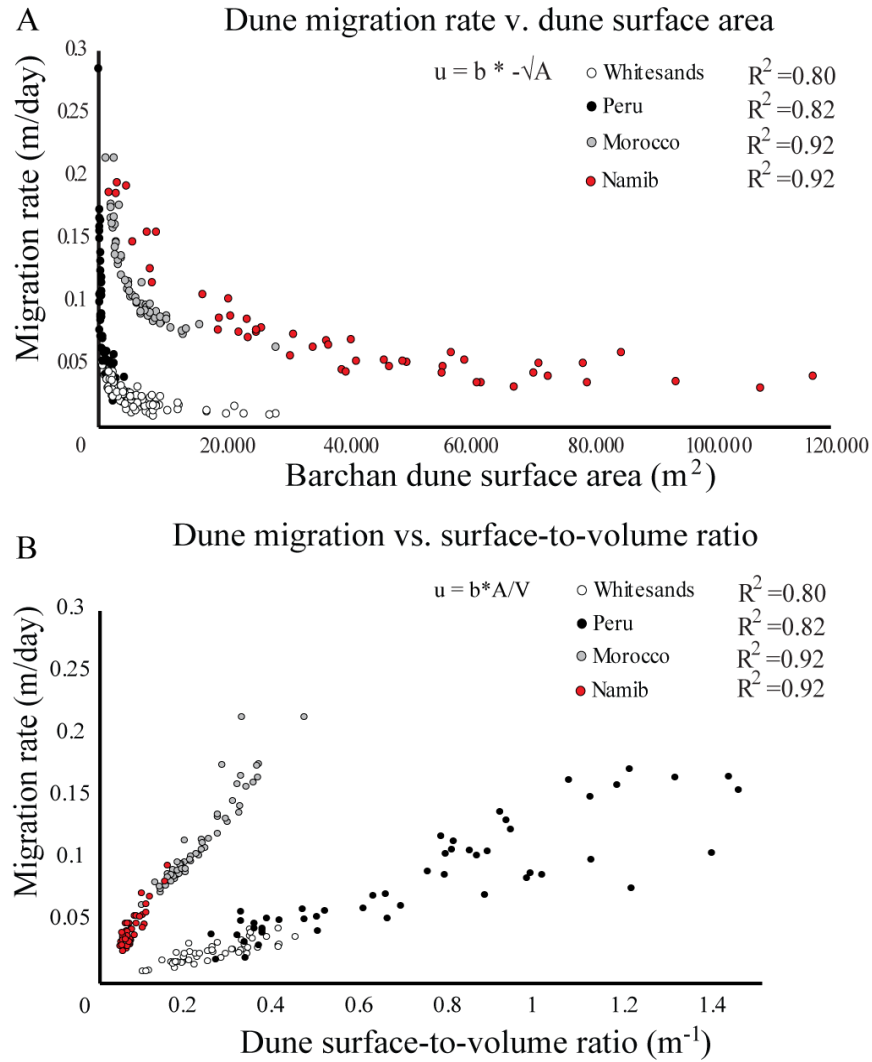

*Figure 5. A. Graph showing dune migration rate versus footprint surface area measured from satellite images of moving barchan dunes.* Each data point represents one dune, each color of dot represents a dunefield. B. The same data as in A. but now the surface-to-volume ratio has been calculated from measurements of stoss side and lee face length (see figure 1). Within each dunefield, surface-to-volume ratios are linearly correlated to migration rate.

The height of each dune is controlled by lee face length and the angle of repose (α), which is almost constant for desert sand under a given gravitational acceleration. On Earth, the angle of repose for typical desert sand is around 34° (Pye and Tsoar, 2008). Accounting for this consistency, surface-to-volume ratios can be calculated using equation (2). To calculate scale-adjusted migration rates, The migration rate of each dune is divided by its surface-to-volume ratio (equation 1). Measured scale-adjusted migration rates are found to be roughly the same and normally distributed within each dunefield but vary by as much as twenty times from one dunefield to another [Footnotes 5, 6].

Now, let's see if the dimensional analysis from the theory is reflected in the real word. The dimension for surface-to-volume ratio is:

$$\frac{m^2}{m^3} = m^{-1}$$

The dimension for scale-adjusted migration rate is:

$$\frac{m}{s} / m^{-1} = \frac{m * m}{s} = \frac{m^2}{s}$$

Scale-adjusted migration rates therefore reflect a surface area covered by a dune per unit of time. This statement is confirmed by measurements, which show that dunes under similar environmental conditions, within the same dunefield, cover the same surface area per unit of time, regardless of their size (supplementary information; *S2*). Plainly, small dunes cover a long, narrow surface area that is equal to that of an adjacent, larger dune's short and wide covered area (Fig. 6).

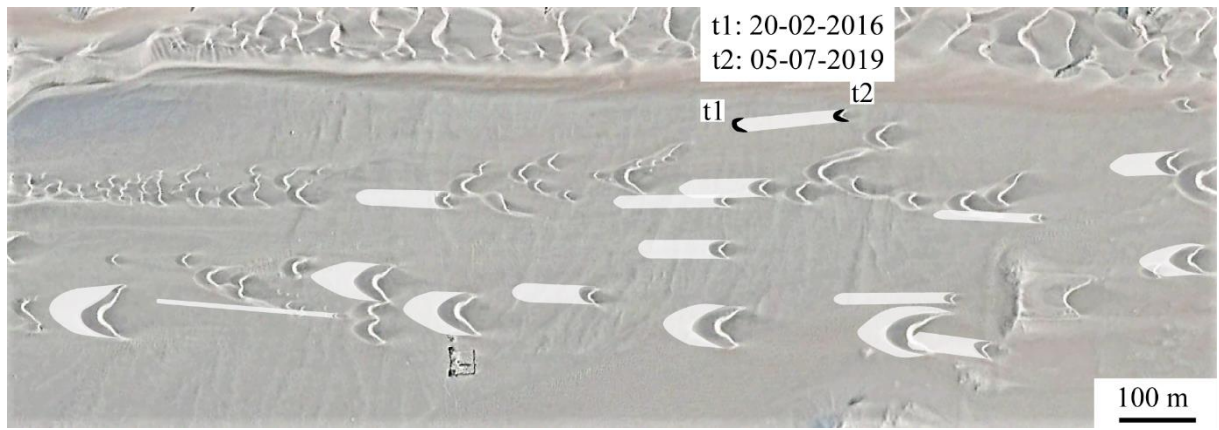

*Figure 6. Satellite image showing that scale-adjusted migration rates are consistent across dunefield. Image shows migrating barchan dunes in Áncash, Peru: 9°20'02"S, 78°20'56"W. Surface areas covered between t1 and t2 are highlighted as grey polygons. In this area, dunes cover $2.1 \pm 1.0$ m²/day which is irrespective of dune size. Surface area covered is equivalent to the scale-adjusted migration rate which is the ratio between dune migration rate and surface-to-volume ratio. As opposed to surface-to-volume ratios, scale-adjusted migration rates depend on environmental factors like windspeed.*

**Parabolic dunes**

Next, it must be shown that variations in the surface-to-volume ratio across a dune's shape define its geometry. Robust evidence for this comes from comparing barchan dunes with parabolic dunes. Parabolic dunes typically form in areas downwind from other dune types, where the water table is closer to the surface. Such conditions cause vegetation to fixate the sediment into place, making parabolic dunes very slow (Reitz et al., 2010). Parabolic dunes are governed by different dynamics than barchan dunes because their surface areas do not promote their motion. Rather, larger vegetated surface areas further inhibit sediment motion for dunes of this type. An ideal area to compare these dynamics is the Whitesands dunefield in New Mexico, USA, where actively migrating barchan dunes and parabolic dunes exist in close proximity (Reitz et al., 2010). Measurements of both dune types where made as per the methodology explained previously and the results are graphed side-by-side in figure (7A). This shows that for parabolic dunes, equation (1) is inversed, with larger parabolic dunes moving faster than smaller ones. Earlier, we hypothesized that for barchan dunes, higher surface-to-volume ratio at the fringes makes them outpace the center, generating the forward-facing 'horned' equilibrium geometry. In the case of parabolic dunes, higher surface-to-volume ratio at the fringes has the inverse effect, causing the center to outpace the fringes and the horns to point backwards. The result of this is that the planform geometries of parabolic dunes are mirrored when compared to barchans, having horns pointed in opposite directions. The same effect is also apparent from measurements of the vertical cross section, which is typically concave for barchans but convex for parabolic dunes (Fig. 7C).

The inverted relationship between size and migration rate for parabolic dunes was also observed in other dunefields like Great Sand Dunes, CO, USA (37°41'N, 105°35'W) and in Saudi Arabia (25°55'N, 49°58'O), see supplementary information *(S3)*. This observation demonstrates that equation (1) applies universally, ubiquitously, and equally *between* dunes and *within* dunes.

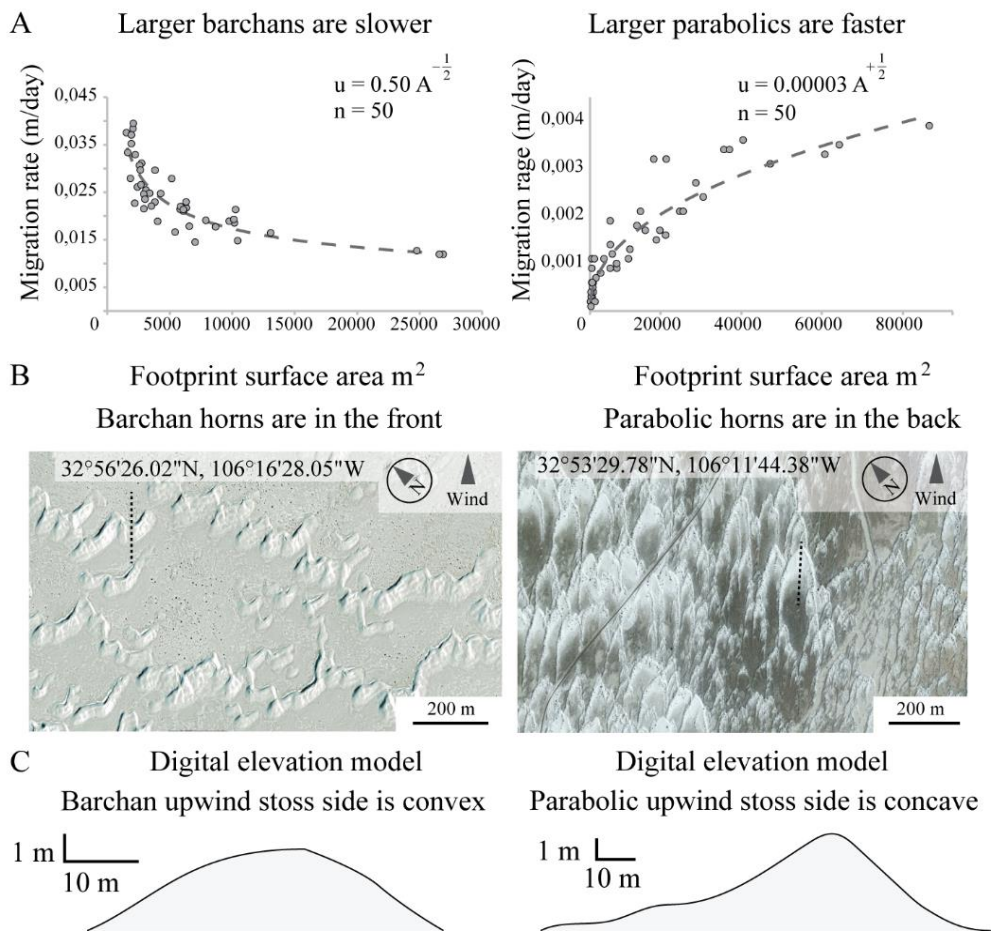

*Figure 7. A: graphs showing the relationship between surface area and migration rate for barchan dunes (left) and parabolic dunes (right) in the same dunefield (Whitesands, NM, USA). Note mirrored relationship between these two. B: Satellite images of barchan dunes and parabolic dunes in Whitesands. C: Vertically exaggerated and smoothed topographic cross sections of a barchan dune and a parabolic dune in Whitesands. Location of cross section in shown by dotted line in (B). 30-cm vertical resolution DEM from LiDAR data is publicly available in Ewing et al. (2012) (35).*

We can now examine the proposed universal relationship between surface-to-volume ratio and equilibrium geometry, by comparing barchan dunes under various environmental conditions (Fig. 8). Very low scale-adjusted migration rates can be measured on Mars and conforming to the model of bed tilting, these barchan dunes have a very low-angle stoss side [Footnote 7]. Contrasting this with dunes in other settings with higher scale-adjusted migration rates, it is found that such dunes universally have a steeper stoss side geometry (Fig. 8). In a sense, a barchan dune is a naturally occurring graph that displays the amount of upward tilting of the sediment under its unique environmental conditions. As predicted by equations

(3 and 4), this concept can be extrapolated across widely different regions and environments, demonstrating that there is a universal relationship between a dune's environment and its geometry. Measured stoss side angles are therefore a predictor for both scale-adjusted migration rate ($R^2 = 0.79$) and sediment transport ($R^2 = 0.69$). Furthermore, there is a strong link between stoss side angle β the difference between b' and Q' ($R^2 = 0.85$).

| Location | Mars | Whitesands | Taklamakan | Peru | Yemen | Egypt | Chad | Libya | Morocco | Namib |
|---|---|---|---|---|---|---|---|---|---|---|
| | 18°14'5"N 100°15'1"O | 32°58'6"N, 106°22'3"W | 36°58'2"N, 82°35'3"E | 09°15'5"S, 078°17'5"W | 14° 3'1"N, 47°49'4"O | 24°55'0"N, 030°28'4"E | 19°20'2"N, 020°17'3"E | 24°24'0"N, 018°27'5"E | 27°16'1"N, 013° 11'2"W | 26°53'3"S, 015°18'6"E |
| b (m²/day) | 0.013 | 0.029 | 0.038 | 0.079 | 0.160 | 0.174 | 0.220 | 0.259 | 0.452 | 0.684 |
| lee/stoss | 0.116 | 0.084 | 0.100 | 0.135 | 0.158 | 0.145 | 0.154 | 0.188 | 0.226 | 0.294 |
| Shape | 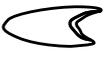 | 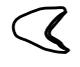 | 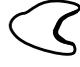 | 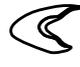 | 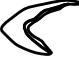 | 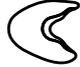 | 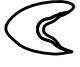 | 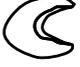 | 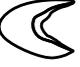 | 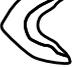 |
| Image (not to scale) | 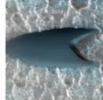 | 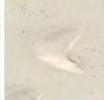 | 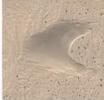 | 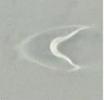 | 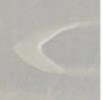 | 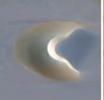 | 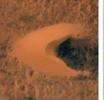 | 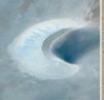 | 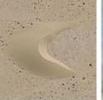 | 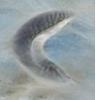 |

*Figure 8. A Compilation of averaged values for scale-adjusted migration rate (b) and lee versus stoss ratios (l/s) in various dunefields. Each measurement is accompanied by a satellite photo and geometric outline of a representative dune within that dunefield. Note that the amount of upward tilting (related to the ratio of lee face length over stoss side length) increases with scale-adjusted migration rate. This relationship can be used to predict scale-adjusted migration rates from satellite photos.*

**Estimating migration rates from dune geometries**

For the final validation, we can combine all aforementioned principles to predict barchan dune migration rates from their shapes and sizes and compare those to measured migration rates. From the theory, it follows that smaller dunes are faster, and steeper dunes have a higher scale-adjusted migration rate. We can infer the relevant quantities from lee face length (l) and stoss side length (s), both of which can easily be measured from satellite images. Assuming a constant angle of repose, the ratio of l and s is proportional to stoss side gradient, and the inverse of length is proportional to surface-to-volume ratio. Consider that migration is inversely proportional to length over constant *'k'*, We rewrite Equation (1.2) to:

$$b = k * \frac{1}{2} \sin \beta$$

This can be combined to Equation (1) to derive dune migration rates:

$$u = k * \frac{1}{2} \sin\beta * \frac{A}{V}$$

The remaining constant value (k) has not been derived in this study, but its quantity can be estimated from the collected data. Predictions from this simple method are statistically significant (k ≈ 3.0e-4 m²/s, $R^2$ = 0.81, P = 5.0e-37, n = 250) but, two important refinements can be made to improve the accuracy.

Firstly, we take note that measured migration rates are time-averaged while migration rate predictions are from a single image and only represent a single moment in time. As wind conditions are dynamic, every dune within a dunefield will exhibit a slightly varying degree of upward tilting compared to adjacent dunes. Owing to equation (1), the geometries of smaller dunes will be affected more by transient changes in the wind so upward tilting at any given moment will be slightly different between dunes of different size, even when the wind is spatially constant across the dunefield. To correct for this, scale-adjusted migration rates can be predicted from stoss side angles values that are averaged across a dunefield.

Secondly, we should consider the differences between the wind-parallel cross section and the complete, three dimensional geometry of a barchan dune. As opposed to linear bedforms, barchan geometries are affected by a growing value for angle β across two directions, namely, they get steeper, and they grow wider (Fig. 4D). Because both the vertical gradient and the aspect ratio (length divided by width) are equally influenced by an increase in β, the extent to which barchan dunes broaden also relies on the scale-adjusted migration rate. To account for this, we can apply the effects of upward tilting quadratically:

$$b = k \left( \frac{l \tan 34}{A'} \right)^2_\mu$$

After estimating scale-adjusted migration rates, the migration rate for each dune can be predicted by applying equation (1):

$$u = k \left(\frac{l \tan \alpha}{A'}\right)_\mu^2 * \frac{A}{V}$$

Similarly to the previous scenario, when estimating the accurate surface-to-volume ratio in three dimensions, surface area expansion caused by an increase in β must also be considered.

$$\frac{A}{V} = \frac{A'}{V'} * \frac{h}{A'} = \frac{A'h}{V'A'} = \frac{h}{V'}$$

$$\frac{h}{V'} = \frac{h}{\frac{1}{2}L * h} = \frac{2}{L}$$

An accurate predictor is therefore:

$$u \approx k \left(\frac{h}{A'}\right)_\mu^2 * \frac{2}{L}$$

In terms of lee face and stoss side length:

$$u = k \left(\frac{l \tan \alpha}{\sqrt{(l \tan \alpha)^2 + s^2}}\right)_\mu^2 * \frac{2}{s+l}$$

The best fit with measured migration rates shows that $k \approx 5.2 * 10^{-4}$ m²/s. Predictions from this methodology resulted in the values shown in Fig. 9B and C which correspond well to measured migration rates ($\alpha = 34°$, $k \approx 5.2 * 10^{-4}$ m²/s, $R^2 = 0.96$, $P = 2.0e\text{-}122$, $n = 250$; see supplementary information; *S4, S5*).

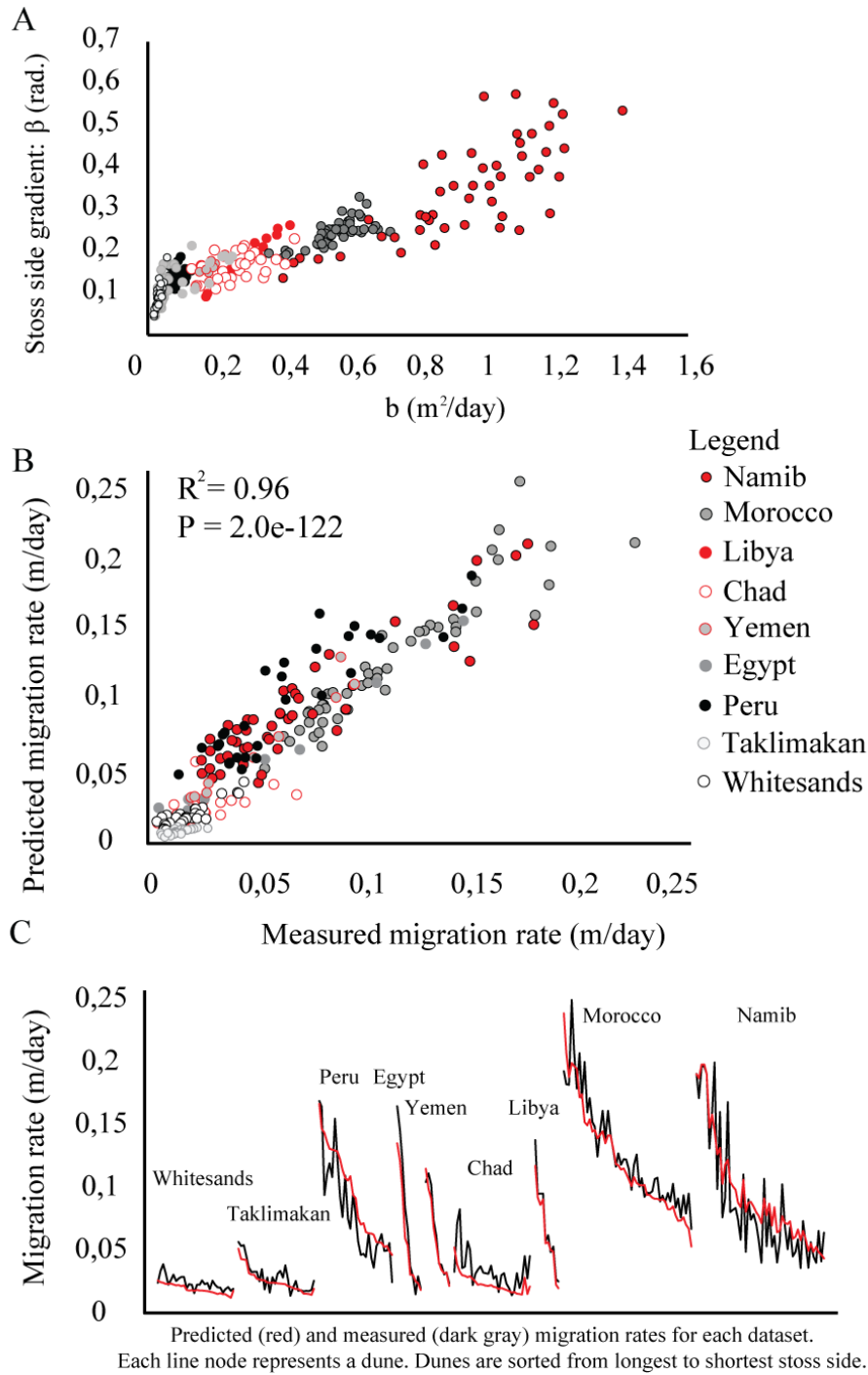

*Figure 9. A. Graph showing the regression analysis of scale-adjusted migration rate (b) and bed shortening (L/A') showing that these parameters are linearly correlated. B: regression analysis indicating that measurements of bed shortening, and surface-to-volume ratio can be used to predict migration rate. C. The same data as in B but graphed differently. predicted (dark grey) and measured (red) migration rates for each dataset. Each line node represents a dune. Dunes are sorted from longest to shortest stoss side. Predictions are accurate across a wide range of dunes and dunefields.*

# DISCUSSION

There are limits to the equilibrium relationship defined in equation (3.2). For example, when stoss side steepness matches the angle of the repose, the sediment on the stoss side will cascade down and the bedform will be eroded. There is therefore a maximum value of sediment tilting under the condition of β = α. Under high current velocity conditions, bedform stability fields will change, eroding ripples or dunes (i.e. 'plane bed', Southard, 1991). From the data collected here, it can be estimated that the upper limit for barchan dune scale-adjusted migration rate is: b ≈ 25 m²/day. However, this estimate does not take into account that wind conditions vary significantly throughout the year, and only a short episode of strong wind may be necessary to erode a dune. Small dunes that adjust their equilibrium geometry rapidly are especially prone to oversteepen and erode during strong gusts or storms. This is reflected in the measurements which show that a dunefield's average dune size correlates with the average scale-adjusted migration rate in that area. The smallest barchan dunes that could be found are in Peru and have low scale-adjusted migration rates of about 1-2 m²/day. In contrast, the Namib dunes may be near the limit of barchan dune dynamics with dunes travelling on average 20 m²/day, up to 25 m²/day. Despite having such widely different environmental conditions, the overall migration rates in these areas are remarkably similar, on average for 0.11 m/day Peru and for 0.12 m/day for Namib, owing to larger dune sizes in the Namib dunefield (Fig. 9C). In fact, the collective measurements from all dunefields demonstrate a weak correlation between windspeed and dune migration rate ($R^2 = 0.27$). This observation may appear counter-intuitive, but follows from equation (4.2), wherein the dune length and dune migration rate are inversely proportional to each other and therefore not directly related to the effective windspeed.

We can extrapolate what happens to β when windspeed increases beyond what is measured in this study by using the relationship:

*Equation* (4.3): $$\beta = atan\left(\left(\frac{k_* \, 2gC\frac{\rho_s}{\rho_w}}{u^3} - \frac{1}{\tan \alpha}\right)^{-1}\right)$$

When this function is graphed, it becomes apparent that beyond the 'oversteepened', 'plane bed' domain of $\alpha > \beta$, there is a domain of stable ($\alpha < \beta$), but negative β. From equation (1.2) we find that a negative value for β implies a negative (i.e. backwards) migration rate. Although unintuitive, this concept is consistent with observations of bedform stability which have backwards migrating antidunes at high current intensities, beyond the 'plane bed' regime (i.e. Harms and Fahnestock, 1965). For aeolian dunes, the wind speeds required to reach this critical threshold are so extreme that sand grains are likely to remain suspended in the air, making aeolian antidunes unlikely to form. However, in different fluid or sedimentological conditions like under water or in fine-grained conditions, this may be more likely. Specifically, a smaller fluid density ratio and a higher value for α would require lower current velocities to enter the negative β domain (Fig. 10). In any case, further testing is needed to prove whether these theoretical relationships are related to real phenomena.

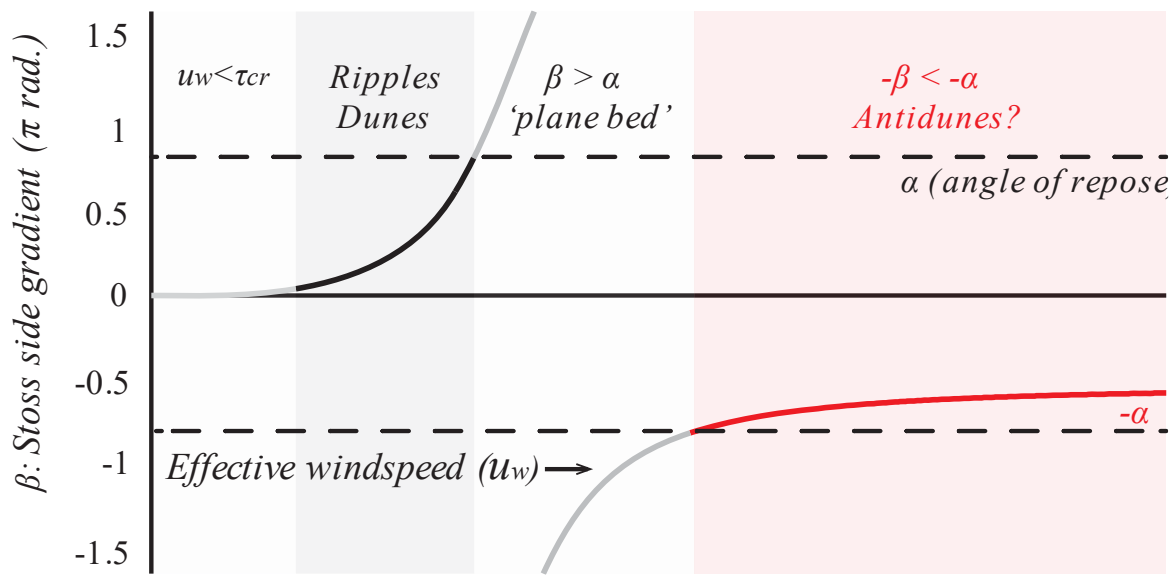

*Figure 10. Extrapolation of Equation (4.3) demonstrates that the wind speed/stoss side identity predicts the bedform stability paradigm. Initially, no bedforms form when the forces are below the critical shear stress threshold. Increasing current velocity generates ripples and eventually (compound) dunes with a higher total stoss side gradient. Bedforms then disappear as the stoss side gradient exceeds the angle of repose, only to reemerge in the negative domain. Following this reemergence, the stoss side points in the opposite direction, resulting in negative migration rates (Equation 1.2).*

The theories presented in this research are applicable only to sediments, which have specific quasi-fluid properties and have an angle of repose. However, some concepts, notably the formation of an angle β from the surface-to-volume ratio axiom, may apply to other fluid -fluid interactions, like wind moving across water or different layers of the atmosphere displacing each other. Since geometry is universal for all these interactions, there may exist a more widely applicable fluid theorem involving the surface-to-volume ratio axiom (Equation 1). Such a theorem may have the potential to simplify existing paradigms of fluid interface phenomena like the Kelvin–Helmholtz instability (Drazin, 2015).

**FOOTNOTES**

Footnote 1: A completely flat bed should be considered to be entirely exposed to the current.

Footnote 2: The stoss side is not diagonal but parabolic, but this simplification does not affect the overall concept. The value for stoss side angle represents the average gradient of the stoss side.

Footnote 3: From equations (1.2 and 3.2) it is already clear that a complete absence of stoss side angle β is only possible when both b' and Q' are zero. To further prove that nonzero bed tilting does now work under the axiom: u = b*A/V, let's consider a nonzero sediment transport:

$$Q = \frac{uV}{L} = u\frac{\frac{1}{2}h*(s+l)}{L} > 0$$

This implies that h, s, l are all nonzero. The condition for zero bed tilting is:

$$\frac{L-A'}{A'} = 0$$

Which implies:

$$L = A'$$

We have already established that for an equilibrium condition: β ≤ α which means that that l > s. We can state that l = s + k so:

$$s + (s + k) = \sqrt{h^2 + s^2}$$

So:

$$(s + (s + k))^2 = h^2 + s^2$$

So:

$$k^2 + 4ks + 4s^2 = h^2 + s^2$$

So:

$$-h^2 + k^2 + 4ks + 3s^2 = 0$$

Since no values are 0 or negative we have reached a contradiction. Some amount of sediment tilting is therefore inevitable when Q > 0.

Footnote 4: A dome-shaped protodune is the most logical shape of initial sediment mount, but a cone-shaped or other type of shaped protodune will have similar surface-to-volume ratio variations across its volume so a similar development will occur. This suggests that regardless of the shape of the initial dune, the combined effects of surface-to-volume ratio variations, and angle of repose will equilibrate in a barchan dune shape when sediment motion occurs. Equation (3) describes the wind-parallel conditions for this equilibrium geometry.

Footnote 5: Calculating scale-adjusted migration rates from a two-dimensional perspective (b; according to equation 2) omits dune widths and therefore results in values that do not reflect the actual values of total surface area covered by the three dimensional entire dune. On average, scale-adjusted migration rates calculated using equation 2 (b') are about 15 times lower than those measured directly from covered surface areas (b; see supplementary information; *S2)*.

Footnote 6: Scale-adjusted migration rates do not significantly deviate from a reference standard deviation. See Kolmogorov–Smirnov normality test in the supplementary information *(S7).*

Footnote 7: Data from the Martian barchans are from the HiRISE orbiter and published in Chojnacki et al. (2020).

# REFERENCES

Liu, L. Y., Skidmore, E., Hasi, E., Wagner, L., & Tatarko, J. (2005). Dune sand transport as influenced by wind directions, speed and frequencies in the Ordos Plateau, China. *Geomorphology*, **67(3-4)**, 283-297.

Ruessink, B. G., & Jeuken, M. C. J. L. (2002). Dunefoot dynamics along the Dutch coast. Earth Surface Processes and Landforms: *The Journal of the British Geomorphological Research* Group, **27(10)**, 1043-1056.

Eastwood, E. N., Kocurek, G., Mohrig, D., & Swanson, T. (2012). Methodology for reconstructing wind direction, wind speed and duration of wind events from aeolian cross-strata. *Journal of Geophysical Research: Earth Surface,* **117(F3).**

Kocurek, G. (1991). Interpretation of ancient eolian sand dunes. *Annual review of Earth and planetary sciences,* **19(1)**, 43-75.

Rodríguez-López, J. P., Clemmensen, L. B., Lancaster, N., Mountney, N. P., & Veiga, G. D. (2014). Archean to Recent aeolian sand systems and their sedimentary record: current understanding and future prospects. *Sedimentology*, **61(6)**, 1487-1534.

Cornish, V. (1897). On the formation of sand-dunes. *The Geographical Journal,* **9(3)**, 278-302.

Drazin, P. G. (2015). Dynamical meteorology| kelvin–helmholtz instability.

Einstein, H. A. (1950). The bed-load function for sediment transportation in open channel flows (No. 1026). US Department of Agriculture.

Harms, J., & Fahnestock, R. K. (1965). Stratification, bed forms, and flow phenomena (with an example from the Rio Grande).

Kennedy, J. F. (1969). The formation of sediment ripples, dunes, and antidunes. *Annual review of fluid mechanics*, **1(1)**, 147-168.

Nishimori, H., & Ouchi, N. (1993). Formation of ripple patterns and dunes by wind-blown sand. *Physical Review Letters,* **71(1)**, 197.

Tsoar, H. (2001). Types of aeolian sand dunes and their formation. In *Geomorphological fluid mechanics* (pp. 403-429). Berlin, Heidelberg: Springer Berlin Heidelberg.

Charru, F., Andreotti, B., & Claudin, P. (2013). Sand ripples and dunes. *Annual Review of Fluid Mechanics*, **45**, 469-493.

Engelund, F., & Fredsoe, J. (1982). Sediment ripples and dunes. *Annual Review of Fluid Mechanics,* **14(1),** 13-37.

Liu, H. K. (1957). Mechanics of sediment-ripple formation. *Journal of the Hydraulics Division*, **83(2)**, 1197-1.

Williams, P. B., & Kemp, P. H. (1971). Initiation of ripples on flat sediment beds. *Journal of the Hydraulics Division*, **97(4),** 505-522.

Andreotti, Bruno, Philippe Claudin, and Stéphane Douady. Selection of dune shapes and Migration rates Part 1: Dynamics of sand, wind and barchans. *The European Physical Journal B-Condensed Matter and Complex Systems* **28.3** 321-339 (2002).

Coco, G., & Murray, A. B. (2007). Patterns in the sand: From forcing templates to self-organization. Geomorphology, **91(3-4)**, 271-290.

Falqués, A., Dodd, N., Garnier, R., Ribas, F., Machardy, L. C., Larroudé, P., ... & Sancho, F. (2008). Rhythmic surf zone bars and morphodynamic self-organization. *Coastal engineering*, **55(7-8)**, 622-641.

Baas, J. H., Best, J. L., & Peakall, J. (2011). Depositional processes, bedform development and hybrid bed formation in rapidly decelerated cohesive (mud–sand) sediment flows. *Sedimentology*, **58(7)**, 1953-1987.

Coleman, S. E., & Eling, B. (2000). Sand wavelets in laminar open-channel flows. *Journal of Hydraulic Research*, **38(5)**, 331-338.

Maun, M. A. (2009). The biology of coastal sand dunes. Oxford University Press.

Schieber, J., & Southard, J. B. (2009). Bedload transport of mud by floccule ripples—Direct observation of ripple migration processes and their implications. *Geology*, **37(6)**, 483-486.

Schieber, J. (2011, May). Shifting paradigms in shale sedimentology-the implications of recent flume studies for interpreting shale fabrics and depositional environments. *In Recovery–CSPG CSEG CWLS Convention* (pp. 1-4).

Yawar, Z., & Schieber, J. (2017). On the origin of silt laminae in laminated shales. *Sedimentary Geology*, **360**, 22-34.

Stokes, G. G. (1850). On the effect of internal friction of fluids on the motion of pendulums. *Trans. Camb. phi1. S0c*, **9(8)**, 106.

Michel, S., Avouac, J. P., Ayoub, F., Ewing, R. C., Vriend, N., & Heggy, E. (2018). Comparing dune migration measured from remote sensing with sand transport prediction based on weather data and model, a test case in Qatar. *Earth and Planetary Science Letters*, **497**, 12-21.

Qian, G., Yang, Z., Tian, M., Dong, Z., Liang, A., & Xing, X. (2021). From dome dune to barchan dune: Airflow structure changes measured with particle image velocimetry in a wind tunnel. *Geomorphology*, **382**, 107681.

Liu, L. Y., Skidmore, E., Hasi, E., Wagner, L., & Tatarko, J. (2005). Dune sand transport as influenced by wind directions, speed and frequencies in the Ordos Plateau, China. *Geomorphology*, **67(3-4)**, 283-297.

Michel et al. (2018)

Baird, T., Bristow, C. S., & Vermeesch, P. (2019). Measuring sand dune migration rates with COSI-Corr and Landsat: *Opportunities and challenges. Remote Sensing,* **11(20)**, 2423.

Bristow, C. S. (2019). Bounding surfaces in a barchan dune: Annual cycles of deposition? Seasonality or erosion by superimposed bedforms?. *Remote Sensing*, **11(8)**, 965.

Bagnold, R. A. (1941). The physics of blown sand and desert dunes. Courier Corporation.

Liu, L. Y., Skidmore, E., Hasi, E., Wagner, L., & Tatarko, J. (2005). Dune sand transport as influenced by wind directions, speed and frequencies in the Ordos Plateau, China. *Geomorphology*, **67(3-4)**, 283-297.

Pye, Kenneth, and Haim Tsoar. Aeolian sand and sand dunes. *Springer Science & Business Media,* **2008**.

Reitz, M. D., Jerolmack, D. J., Ewing, R. C., & Martin, R. L. (2010). Barchan-parabolic dune pattern transition from vegetation stability threshold. *Geophysical Research Letters*, **37(19).**

Ewing, Kocurek, Jerolmack, Bustos, 2012, *Constructing the Dune-Field Pattern at White Sands National Monument.* White Sands Science Symposium, Research brief.

Sauermann, G., Rognon, P., Poliakov, A., & Herrmann, H. J. (2000). The shape of the barchan dunes of Southern Morocco. *Geomorphology*, **36(1-2)**, 47-62.

Wayne, D. (1990). Kolmogorov–Smirnov one-sample test. Applied *nonparametric statistics*, **2.**

Chojnacki, M., Fenton, L. K., Weintraub, A. R., Edgar, L. A., Jodhpurkar, M. J., & Edwards, C. S. (2020). Ancient Martian aeolian sand dune deposits recorded in the stratigraphy of Valles Marineris and implications for past climates. *Journal of Geophysical Research: Planets*, **125(9),** e2020JE006510.

# SUPPLEMENTARY INFORMATION

S1: Dunefield measurements

S2: Surface areas covered

S3: Parabolic dunes

S4: Migration rate prediction

S5: Linear Regression

S6: Whitesands

S7: Normality test

# ACKNOWLEDGEMENTS

I would like to express my sincere appreciation for the wealth of invaluable comments and insights provided by Hanaga Simabrata, Joost Bergen, Manolo Rodriguez, and Gunther Cornelissen.